\documentclass[a4paper, conference]{IEEEtran}
\IEEEoverridecommandlockouts
\usepackage{amsmath,amssymb,amsfonts}
\usepackage{algorithmic}
\usepackage{graphicx}
\usepackage{textcomp}
\usepackage{xcolor}
\usepackage[square,numbers,sort]{natbib}
\usepackage[obeyspaces]{xurl}
\usepackage[T1]{fontenc}
\usepackage{tikz}
\usetikzlibrary{external}
\usepackage{multirow}
\usepackage{enumitem}
\usepackage[left=1.62cm,right=1.62cm,top=2.4cm, bottom=4.3cm]{geometry}

\usepackage[compact]{titlesec}
\titlespacing{\section}{0pt}{5pt}{2pt}
\titlespacing{\subsection}{0pt}{2pt}{2pt}
\titlespacing{\subsubsection}{0pt}{2pt}{2pt}
\begin{document}

\title{BinImg2Vec: Augmenting Malware Binary Image Classification with Data2Vec\\
{\footnotesize \textsuperscript{}}
\thanks{}
\vspace{-30pt}

}

\author{\IEEEauthorblockN{Lee Joon Sern}
\IEEEauthorblockA{\textit{Ensign Labs} \\
\textit{Ensign InfoSecurity}\\
Singapore \\
lee\_joonsern@ensigninfosecurity.com}
\and
\IEEEauthorblockN{Tay Kai Keng}
\IEEEauthorblockA{\textit{Ensign Consulting} \\
\textit{Ensign InfoSecurity}\\
Singapore \\
kaikeng\_tay@ensigninfosecurity.com}
\and
\IEEEauthorblockN{Chua Zong Fu}
\IEEEauthorblockA{\textit{Ensign Consulting} \\
\textit{Ensign InfoSecurity}\\
Singapore \\
zongfu\_chua@ensigninfosecurity.com}
}

\IEEEaftertitletext{\vspace{-30pt}}
\maketitle

\begin{abstract}
    Rapid digitalisation spurred by the Covid-19 pandemic has resulted in more cyber crime. Malware-as-a-service is now a booming business for cyber criminals. With the surge in malware activities, it is vital for cyber defenders to understand more about the malware samples they have at hand as such information can greatly influence their next course of actions during a breach. Recently, researchers have shown how malware family classification can be done by first converting malware binaries into grayscale images and then passing them through neural networks for classification. However, most work focus on studying the impact of different neural network architectures on classification performance. In the last year, researchers have shown that augmenting supervised learning with self-supervised learning can improve performance. Even more recently, Data2Vec was proposed as a modality agnostic self-supervised framework to train neural networks. In this paper, we present BinImg2Vec, a framework of training malware binary image classifiers that incorporates both self-supervised learning and supervised learning to produce a model that consistently outperforms one trained only via supervised learning. We were able to achieve a 4\% improvement in classification performance and a 0.5\% reduction in performance variance over multiple runs. We also show how our framework produces embeddings that can be well clustered, facilitating model explanability.
\end{abstract}


\section{Introduction}
    The rapid digital adoption, spurred largely by the Covid-19 pandemic, to support work-from-home arrangements all over the world, has resulted in a surge in worldwide Internet usage. \cite{internetPandemic}. This has in turn fueled cyber-attacks; In 2021, Governments worldwide experienced a 1,885\% increase in ransomware attacks, while the healthcare industry alone experienced a 775\% increase in those attacks in 2021 \cite{surgeattacks}. In fact, Malware-as-a-service (MaaS) is now a booming business for cybercrime organisations with new malware samples increasing 106\% year-on-year in 2021 \cite{maas}. Traditional signature-based matching approaches to detect and profile malware into their respective families may no longer work simply because of the inability to capture and obtain signatures for every single malware out there in the wild. In fact, \citeauthor{evaded} found that around three quarters of malware samples collected in Q1 2021 would have evaded traditional signature-based secturity tools \cite{evaded}. As such, it is increasingly important for cyber defenders to move away from signature-based methods and find generalizable, scalable methods to detect and profile the rapidly increasing number of malware. In this paper, we study how image classification methods can be applied to the malware family classification problem. This is particularly important as the ability to classify the malware family will provide cyber defenders an idea of what the malware was designed to do, thus reducing the search space for cyber incident responders when trying to contain a breach by the malware.
    
\section{Related Work}

    In 2011, \citeauthor{firstmalimg} coined a novel idea to turn malware binaries into byte images. They then made use of the well known GIST image descriptor method to extract feature vectors from each of the images. They found that malware belonging to the same family tend to have similar feature vectors resulting in them being loosely clustered together \cite{firstmalimg}. Though extremely promising, it was noted that improvements could be made through a learning based algorithm such as artificial neural networks\cite{ann2015}. 
    
    In 2015, \citeauthor{ann2015} proposed to enhance \citeauthor{firstmalimg}'s work by applying the Discrete Wavelet Transformation on the malware images to extract additional features, on top of GIST, before passing them all through a multi-layered perceptron (MLP) neural network to train a malware family classifier \cite{ann2015}. Subsequently, with the advent and adoption of Convolutional Neural Networks (CNN) over multiple domains, \citeauthor{cnn2017} proposed passing the entire malware image binary through a CNN without any hand engineered features \cite{cnn2017}. They showed that they were able to get 93.17\% classification accuracy over 32 malware families. Separately, \citeauthor{cnn2020} also showed good performance of CNN based techniques, achieving 99.24\% accuracy over 25 families \cite{cnn2020}. With deep learning showing promising results in this domain, \citeauthor{cnn2021}, in 2021, conducted an empirical analysis of various well known deep learning methods ranging from MLPs, CNNs, Recurrent Neural Networks (RNN), Long Short-Term Memory (LSTM), etc. He found that CNN based methods provided the best results empirically \cite{cnn2021}.
    \begin{figure*}[!htp]
        \centering
        \includegraphics[width=15cm, height=4cm]{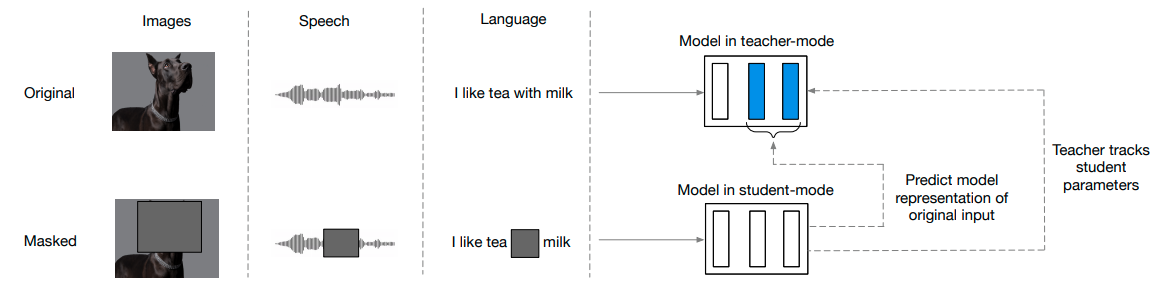}
        \caption{Data2Vec Framework}
        \label{fig:data2vec}
    \end{figure*}
    
    To the best of our knowledge, all previous works in this domain trained the deep neural networks via supervised learning, using the softmax cross-entropy loss function, which is commonly used to train multi-class classifiers. While supervised learning has conventionally been shown to be the de-facto method to train neural networks, if labels are available, due to the fact that its associated loss functions like the softmax cross-entropy loss function is convex, a number of recent works have proposed methods that have been shown to be able to surpass supervised-only training. In particular, recent work showed that augmenting supervised learning with self-supervised learning techniques could improve model performance significantly. For example, \citeauthor{maskedautoenc}, in 2021, showed that randomly masking pixels of an input image helps an autoencoder learn more robust embeddings, which would be useful for subsequent fine tuning tasks \cite{maskedautoenc}. Separately, \citeauthor{beit} showed that pre-training transformer based neural networks via a self-supervised approach helped improve classification accuracy by approximately 2\% \cite{beit}. A significant drawback of these findings is that they are domain dependent, meaning that some of these proposed methods only work well for natural language processing, while others work well on image tasks; they may not work well across multiple domains (i.e. modalities). To address this issue \citeauthor{data2vec}, in Feb 2022, proposed a possible modality agnostic framework that they showed could work equally well across multiple modalities, from image processing to natural language processing and even speech processing \cite{data2vec}.

    In particular, \citeauthor{data2vec} proposed the data2vec framework as shown in Fig. \ref{fig:data2vec} \cite{data2vec}. Similar to \citeauthor{maskedautoenc}, they first proposed to apply a random mask to the input, be it an image, speech or even text. The unmasked input is then passed to a teacher neural network (TNN) to output a series of embeddings. Similarly, the masked input is passed to a student neural network (SNN) to output another series of embeddings. It is important to note that the TNN and the SNN are identical in architecture. However, the weights of the TNN are actually an exponential moving average of the SNN as it is updated. The weights of the TNN are parameterised as shown in Equation \eqref{teacher_update}, where $\triangle_{n+1}$ represents the weights of the TNN at training step $n+1$, $\triangle_n$ represents the weights of the TNN at step $n$, $\theta$ represents the weights of the SNN at step $n$ and $\tau$ represents the exponetial moving average multiplier. Finally, the loss between the series of embeddings output by both the TNN and SNN is minimised, while holding the embedding output of the TNN constant (i.e. a fixed non-differentiable target).
    
    \small
    \begin{equation}
        \triangle_{n+1} = \triangle_n \tau + (1-\tau)\theta_n
        \label{teacher_update}
    \end{equation}
    \normalsize
    
    Data2Vec has been well received by the deep learning community not only because of it being modality agnostic but also because it precludes the need to train the decoding portion of auto-encoders. Conventionally, one would train an auto-encoder neural network comprising an encoder and a decoder, to obtain embeddings that contain summarized information of the encoder's input \cite{maskedautoenc}. However, the decoder is rarely used once the embeddings are obtained as the main purpose of the autoencoder was to obtain the encoder and its ouput embeddings. Furthermore, conducting backpropagation of gradients through the decoder is an expensive process as the decoder needs to reconstruct the encoder's input. Data2Vec is a significant step up as it precludes the need to train a decoder. Instead, it bootstraps the training using an exponential moving average of the SNN's weights. This together with its modality agnostic characteristic have resulted in data2vec garnering alot of interest from the research community recently \cite{d2v1} \cite{d2v2}.
    
    We now highlight our main contributions. Our major contribution is the study of how the data2vec framework can be applied to the malware image binary classification problem. We will validate if data2vec is applicable even to malware binary images and we will also propose a method to incorporate the data2vec framework in order to train a malware binary image family classifier in an end-to-end fashion. We will also evaluate the performance of our proposed framework. Our other contribution is an in-depth study of the neural network's output to better understand what the network has learnt in order to aid explanability and deeper analysis.

    \begin{figure*}[!htp]
        \centering
        \includegraphics[width=11cm, height=10cm]{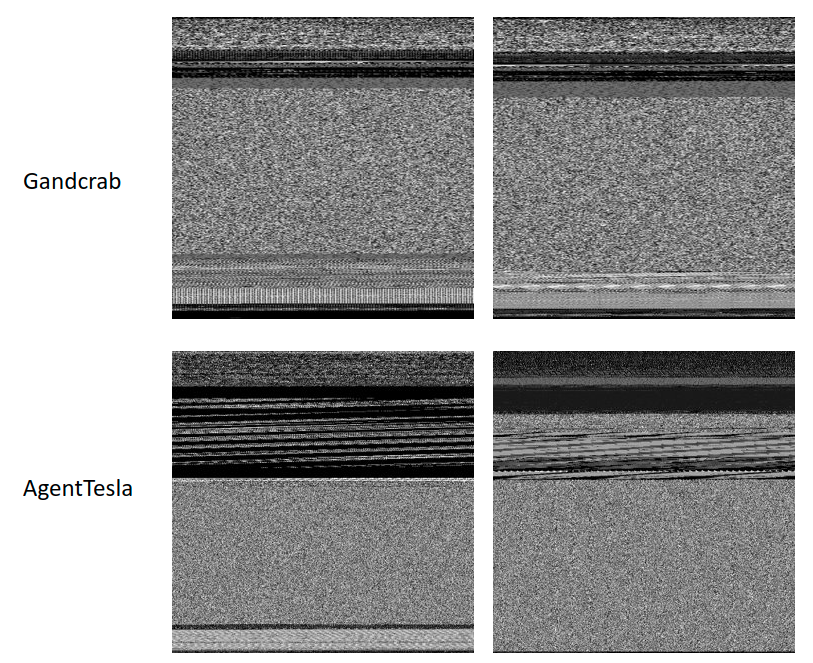}
        \caption{Examples of Malware Binary Images}
        \label{fig:malbinimg}
    \end{figure*}
    
    \begin{figure*}[!htp]
        \centering
        \includegraphics[width=12cm, height=3cm]{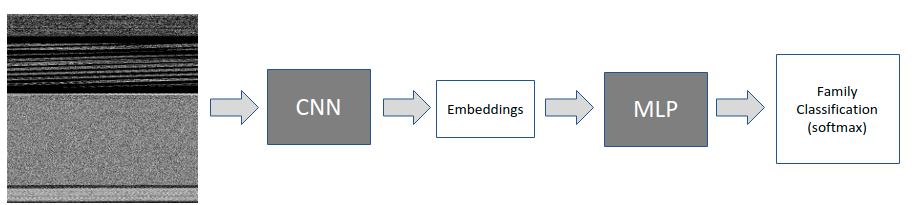}
        \caption{Model Design}
        \label{fig:model_design}
    \end{figure*}
\section{Preliminaries and Problem Setup}
    In this section, we will review how malware binaries can be converted to images and formally define the classification problem. In this paper, the malware binaries that we analyse pertain only to Windows malware. Thus, they would be in Portable Executable format, having file extensions .bin, .dll and .exe \cite{cnn2020}. These malware binaries comprise of a series of bits and can be converted 8 bits at a time to pixels in a grayscale image, comprising of textural patterns. This is because each byte (i.e. 8 bit) can be mapped to an integer between 0 to 255 and the entire byte array can be mapped to a grayscale image.
    
    Our dataset of malware binaries were obtained from multiple online data sources \cite{malsamp1} \cite{malsamp2} and comprise of 61 Windows malware families. To convert them to images, we first studied the size of the byte array. We found that a majority of the malware binaries, when converted to a byte array had lengths around 640,000 (i.e. 640 kilobytes). Thus we set the width of the image to be 800 so that most of the images would be square in nature. This would facilitate downstream resizing without significant information loss. It would also facilitate easier CNN designs as the kernel sizes can all be square too. The steps below summarize how we preprocess each of the malware binaries into images before passing them as input to the classification model.
    \begin{enumerate}
        \item Read in malware binaries as a byte array.
        \item Zero pad the byte array so that its length is divisible by 800.
        \item Reshape the zero padded byte array such that its width is 800.
        \item Resize the entire array into a 400 $\times$ 400 grayscale image. This facilitates a larger batch size when training on RAM constrained Graph Processing Units (GPUs)
    \end{enumerate}
    
    Figure \ref{fig:malbinimg} depicts samples of 2 malware families from our dataset after the above preprocessing steps. As can be seen, malware of the same family show similar visual textures. In particular, those of the Gandcrab malware have near identical characteristics at the top of the image, while those of AgentTesla have diagonal lines cutting across the image. These hint at the possibility of image classification methods being used to classify malwares into their family.
    After images of the malware binaries are obtained, methods described in previous works \cite{cnn2017}\cite{cnn2020}\cite{cnn2021} all train a CNN using the softmax cross entropy loss function. The softmax cross entropy loss is as defined in \eqref{ce_loss}, where $m$ is the batch size, $y_i$ is a one hot vector indicating the labelled class for that particular sample and $\hat{y_{i}}$ is the softmax output of the CNN, which is actually a discrete probability distribution describing the probability a particular malware binary image sample belongs to each of the known malware families in the dataset. This loss function is convex and is therefore typically used to train multi-class classifiers.
    
    \small
    \begin{equation}
        L_{ce} = -1/m \sum_{i=1}^{m} y_{i} \cdot ln(\hat{y_{i}}) \label{ce_loss}
    \end{equation}
    \normalsize
    
    As mentioned earlier, recent deep learning research has shown that incorporating self-supervised learning as part of the neural network's training process can improve model performance \cite{maskedautoenc}\cite{beit}\cite{data2vec}. Inspired by these promising results, we will next describe how we incorporated \citeauthor{data2vec}'s data2vec algorithm into our training process.
    
\section{Methodology}
    One of the drawbacks of neural networks is the lack of explanability. To the best of our knowledge, all previous works in this domain treated the CNN models that they trained as a blackbox. After training the CNN, they only evaluated the output of the CNN with no additional insight as to how the model was able to produce its prediction. In this paper, we propose to make use of data2vec to not only enhance performance but also improve explanability of the neural network so as to provide greater confidence that the model is indeed learning as we expect it to be. To do this, we broke the model into 2 sub nueral networks as shown in Figure \ref{fig:model_design}. Essentially, we will make use of a CNN to ingest the various malware binary images and produce embeddings. These embeddings will then be passed to a MLP to produce a softmax output, indicating the probability of the input malware binary image belonging to each of the known malware families in the training dataset. Intuitively, we would like the CNN to be a feature extractor while the MLP to be a feature analyzer and class prediction module.

\begin{figure*}[!htp]
        \centering
        \includegraphics[width=13cm, height=6cm]{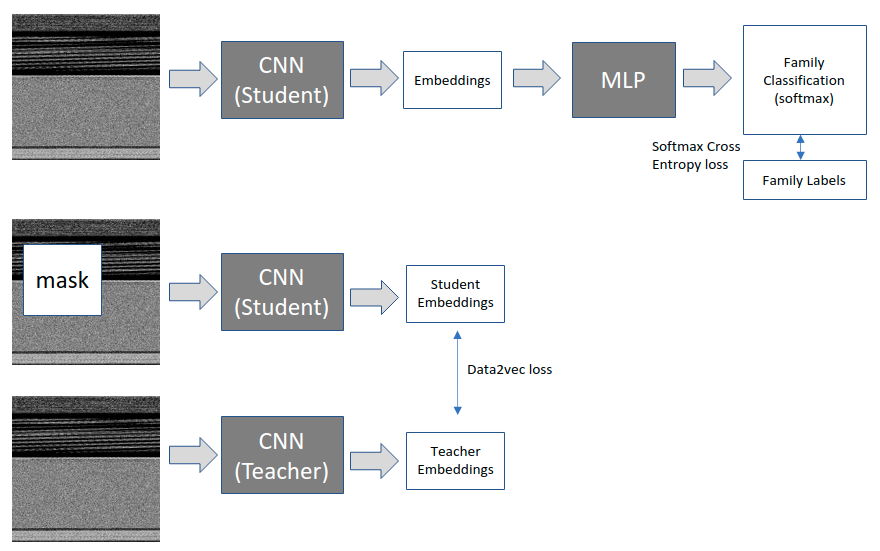}
        \caption{BinImg2Vec Framework}
        \label{fig:workflow}
    \end{figure*}

    We propose to train the entire network (both the CNN and MLP) using the softmax cross entropy loss, while regularising the embeddings output by the CNN using the data2vec framework. Figure \ref{fig:workflow} shows our proposed framework. 
    
    As shown, it comprises 2 main workflows. The first workflow is the conventional supervised learning workflow whereby the entire model, comprising both the CNN and MLP are trained end-to-end using the labelled dataset via the conventional softmax cross entropy loss function described in \eqref{ce_loss}. Do note that in this workflow, the entire unmasked image is fed to the CNN model. Also, we use the student CNN for this workflow as the teacher CNN is in fact a frozen moving average of the student CNN (i.e. gradient propagation should not go through the teacher CNN). The second workflow aims to regularise the output of the student CNN to encourage the model's output embeddings contain information representative of the input image. It also helps to prevent the model from over-fitting to the dataset. This workflow is as what was proposed in \cite{data2vec} whereby the student CNN is input a masked version of the malware binary image and the teacher CNN is input the entire malware binary image. The difference between the output embeddings are then minimized via \eqref{d2vloss}, where $z_t$ is the output embeddings produced by the teacher CNN, $\hat{z_t}$ is the output embeddings produced by the student CNN. Note that $\beta$ is used to control the loss function's sensitivity to outliers. In our experiments, we set this to 0.5.
    
    Our final proposed composite loss function to train the model end-to-end is \eqref{comp_loss}, where $\lambda$ is a weighting factor, which we set to 1.
    
    \small
    \begin{equation}
        L_{data2vec}= 
    \begin{cases}
        \frac{1}{2} (z_t-\hat{z_t})^2 / \beta,& \text{if } |z_t-\hat{z_t}|\leq \beta\\
        (|z_t - \hat{z_t}| -\frac{1}{2} \beta ),              & \text{otherwise} 
    \end{cases}\label{d2vloss}
    \end{equation}
    \normalsize
    
    \small
    \begin{equation}
        L = L_{ce} + \lambda L_{data2vec} \label{comp_loss}
    \end{equation}
    \normalsize
    
    \subsection{Network Design}
    Table \ref{tab:network_archi} summarizes the network architectural design. It should be noted that although \citeauthor{data2vec} proposed a transformer architecture, he also noted that the data2vec methodology would also be applicable to other alternative architectures \cite{data2vec}. In this paper, we chose a CNN architecture in place of transformer given that it has achieved good results in this domain and is less computationally intensive. The first 3 convolutional layers that does "same" padding aims to extract features from the input image, with minimal information loss from the periphereal pixels. The next few convolutional layers are feature extractors. It should be noted that the output of the convolutional layers is a matrix of size 3 $\times$ 3 $\times$ 256. This is then passed through a dense layer (i.e. Dense (CNN)) to output a matrix of the same size but with linear activation. This matrix forms the embedding layer, which will be trained using $L_{data2vec}$. The next few layers are part of the MLP network. It first flattens the 3 dimensional matrix into a single vector, before passing it through 3 Dense layers of the same number of units. These Dense layers follow the ResNet architecture to facilitate back-propagation of gradients. In addition, we added a dropout of 0.2 throughout the entire network to prevent overfitting.
    
    \begin{table}[!htbp]
    \caption{Neural Network Architecture}
    \label{tab:network_archi}
    \centerline{
        \begin{tabular}{|p{0.7cm}|p{0.7cm}|p{0.7cm}|p{1.8cm}|p{0.9cm}|p{1.3cm}|}
        \hline
        \textbf{Filters} & \textbf{Stride} & \textbf{Kernel} & \textbf{Convolution Type} & \textbf{Padding} & \textbf{activation} \\
        \hline
        32 & [1,1] & [3,3] & Conv2D (CNN) & Same & leaky relu\\
        64 & [1,1] & [3,3] & Conv2D (CNN) & Same & leaky relu\\
        128 & [1,1] & [3,3] & Conv2D (CNN) & Same & leaky relu\\
        
        128 & [2,2] & [5,5] & Conv2D (CNN) & Valid & leaky relu \\
        128 & [2,2] & [5,5] & Conv2D (CNN) & Valid & leaky relu\\
        256 & [2,2] & [5,5] & Conv2D (CNN) & Valid & leaky relu \\
        256 & [2,2] & [5,5] & Conv2D (CNN) & Valid & leaky relu\\
        256 & [2,2] & [5,5] & Conv2D (CNN) & Valid & leaky relu\\
        256 & [2,2] & [5,5] & Conv2D (CNN) & Valid & leaky relu\\
        
        256 & - & - & Dense (CNN) & - & linear\\
        - & - & - & Flatten (MLP) & - & -\\
        128 & - & - & Dense (MLP) & - & leaky relu\\
        128 & - & - & Dense (MLP) & - & leaky relu\\
        128 & - & - & Dense (MLP) & - & leaky relu\\
        61 & - & - & Dense (MLP) & - & softmax\\

        \hline
        \end{tabular}
        }
    \end{table}
    
    \subsection{Experimental Setup}
    To showcase the improvement of our methods over the existing literature, we will train 2 models on 90\% of our dataset, segmented by family to ensure that the test dataset has every malware family. The first model is trained using only $L_{ce}$ while the second model is trained using our composite loss function $L$. We will then repeat this process thrice using different random seeds to objectively determine whether our proposed framework can improve the malware family classification accuracy. All models took approximately 12 hours of training to converge. Loss curves were also observed throughout the entire training process to ensure convergence has been reached.
    
\section{Results and Discussion}

    The results of our experiments are as shown in \ref{tab:classification_accuracy}. As can be seen our proposed method of training the malware binary image classification model produce a marked improvement of nearly 4\% consistently across the 3 runs. Furthermore the variance across the runs was reduced from 0.7\% to 0.2\%. This shows that our framework of using both supervised and self-supervised training to train the neural network gives better results.
    
    \begin{table}[htbp]
    \caption{Accuracy \& F1-Score of various Experiments and Scenarios}
    \label{tab:classification_accuracy}
    \centerline{
        \begin{tabular}{|p{2cm}|c|c|c|c|}
        \hline
        \textbf{ } & \textbf{Experiment} & \textbf{$L_{ce}$} & \textbf{$L$} \\
        \hline
        \multirow{3}{2cm}{Family Classification Accuracy} & 1 & 0.956 & 0.989\\
                 & 2 & 0.950 & 0.991 \\ 
                 & 3 & 0.949 & 0.990 \\ 
        \hline
        Mean Accuracy & & 0.952 & 0.991 \\
        \hline
        
        \end{tabular}
        }
    \end{table}
    \begin{figure*}[!htp]
        \centering
        \includegraphics[width=14cm, height=6cm]{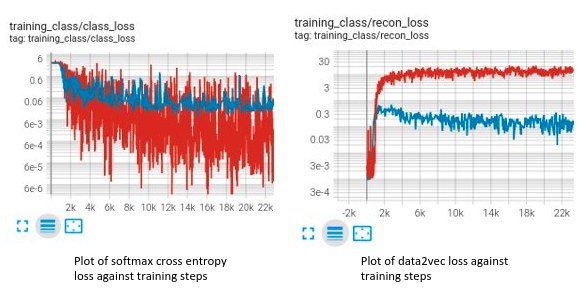}
        \caption{Comparing Training Loss Curves}
        \label{fig:loss_curves}
    \end{figure*}
    
    \begin{figure*}[!htp]
        \centering
        \includegraphics[width=15cm, height=7cm]{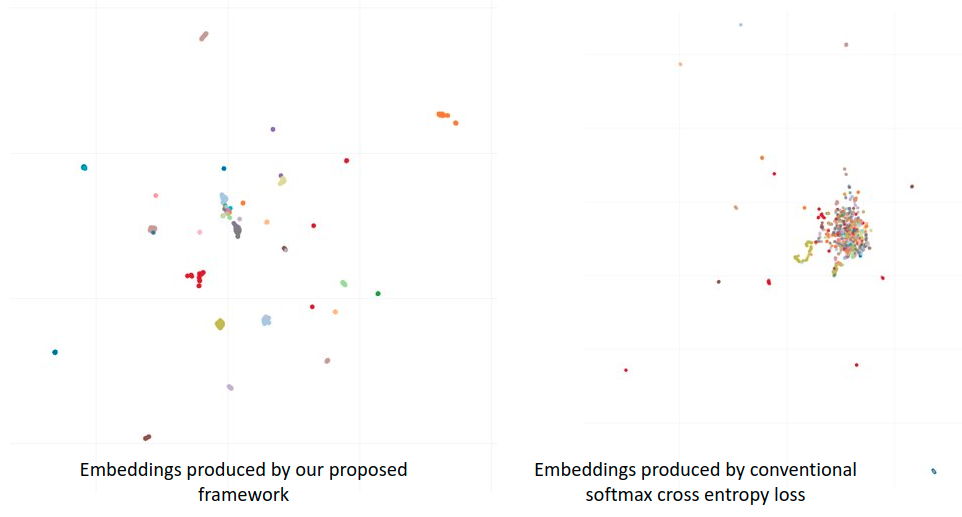}
        \caption{Comparison of Embeddings}
        \label{fig:embeddings_both}
    \end{figure*}
    To better understand why this is so, we present the training loss curves of the 2 different approaches in Figure \ref{fig:loss_curves}. The red curve depicts the softmax cross entropy loss as the model is trained using just $L_{ce}$ while the blue curve depicts the softmax cross entropy loss component of our composite loss function when the model is trained using $L$. As can be seen, the blue curve settles on a higher loss but is much more stable compared to the red one. We opine that this is one reason why the model, when trained using $L$ results in less variation over multiple seeds compared to if it were just trained using $L_{ce}$.

    Secondly, we also note that the $L_{data2vec}$ loss function for the red curve settled on at least 2 orders of magnitude higher than the blue curve. This is as expected as the red curve was not trained using this loss function (although we log it too for comparison). However, it is interesting to discover that the difference would be so huge. Intuitively, what this means is that the embeddings produced by the 2 networks could have vastly different meanings. For the blue curve, the embeddings must contain information of the original input image. On the other hand, this may not be the case for the red curve; its embeddings may well be tailored to overfit to the training dataset.

    We attempt to explain this phenomenon by trying to understand the inner-workings of the model. To do this, we took the embeddings produced by the CNN model (1 trained using $L_{ce}$ and another using $L$) and passed them through UMAP \cite{umap} to reduce the embeddings to 2 dimensions, allowing us to plot them in a 2-dimensional plane. As can be seen in Figure \ref{fig:embeddings_both}, the embeddings produced by our proposed framework are actually well clustered by the various families in the dataset (i.e. each color represents a particular family and each cluster is predominantly made up of one color, indicating that each cluster is predominantly made up of 1 family), even though this was not explicitly expressed in the $L_{data2vec}$ loss function. On the other hand, the embeddings of the model trained using just the softmax cross entropy loss is rather erratic with no clear cluster. This characteristic is extremely helpful to us because of the following:

    \begin{enumerate}
        \item Firstly, it enhances the explanability of the model. We are now sure that the CNN trained under our proposed framework indeed extracts interpretable features for the MLP network to analyse. In contrast, this is not the case if the model was simply trained using the $L_{ce}$ loss function.
        \item Secondly, it facilitates the discovery of new malware families. If the MLP's confidence is low, we can analyse the embeddings output by the CNN to see if it is indeed far from all the known clusters. If so, the malware sample is probably a new malware family.
    \end{enumerate}

\section{Conclusion and Future Work}
    In this paper we have made a few findings. First, we verify that the data2vec framework indeed also works in the cyber domain. To the best of our knowledge, we are the first to apply this framework to the cybersecurity domain. Secondly, we demonstrate how the addition of $L_{data2vec}$ as a regularising term in the loss function is able to improve malware family classification and also stabilise training so that the variance of the model's performance over multiple runs is relatively stable. Thirdly, we show how our proposed framework can enhance explanability of the model and possibly identify new malware families through clustering of the embeddings output by the CNN network. In this work, we have verified our proposed framework over 61 malware families, which is more than all previous work to the best of our knowledge. Future work, could extend this to even more families and also explore malware functionality classification (i.e. Trojan, Worm, etc.)

\bibliographystyle{unsrtnat}
\bibliography{references}

\end{document}